\documentclass[twocolumn,showpacs,preprintnumbers,amsmath,amssymb,aps]{revtex4-1}
\usepackage{graphicx}
\usepackage{dcolumn}
\usepackage{bm}
\usepackage{float}
\usepackage{soul}

\begin{document}

\title{Dark state population determines magnetic sensitivity in radical pair magnetoreception model}

\author{Bao-Ming Xu$^{1}$}
\author{Jian Zou$^{2}$}%
\email{zoujian@bit.edu.cn}

\affiliation{$^{1}$School of Physics, Qufu Normal University, Qufu 273165, China}
\affiliation{$^{2}$School of Physics, Beijing Institute of Technology, Beijing 100081, China}%


\begin{abstract}
What is the real role of the quantum coherence and entanglement in the radical pair (RP) compass, and what determines the singlet yield have not been fully understood. In this paper, we find that the dark states of the two-electron Zeeman energy operator (TEZE) play an important role in the RP compass. We respectively calculate the singlet yields for two initial states in this dark state basis: the coherent state and the same state just removing the dark state coherence. For the later there is neither dark state coherence nor entanglement in the whole dynamical process. Surprisingly we find that in both cases the singlet yields are the same, and based on this result, we believe that the dark state population determines the singlet yield completely, and the dark state coherence and entanglement have little contribution to it. Finally, we also find that the dark state population as well as the singlet yield anisotropy is fragile to the vertical magnetic noise. However, the orientation is robust and is even enhanced by the parallel magnetic noise because the dark states expand a decoherence-free subspace. The dark state population as well as the orientation is more robust to the hyperfine coupling noise.
\end{abstract}


\maketitle

\section{Introduction}.
It is well known that certain migratory birds can use the Earth's magnetic field for orientation and navigation. As one of the main hypotheses to explain the magnetic sensing, the RP mechanism \cite{Ritz2000,Steiner,Wiltschko,Maeda,Rodgers,Kominis,Gauger,Jayendra,Cai1,Cai2,Cai3, Cai4,Hannah,Xu} was first proposed in the pioneering work by Schulten \emph{et al}. \cite{Schulten}. In the RP mechanism, the spin relaxation should be slow enough, i.e., the lifetime should be long enough, generally in the order of $10^{-6}$-$10^{-5}$s \cite{Rodgers,Jayendra}, or even $10^{-4}$s \cite{Gauger}. Several important experiments support this RP mechanism \cite{Wiltschko72,Wiltschko78,Wiltschko06,Wiltschko13,Ritz04,Ritz05,Ritz09,Wiltschko15,Wiltschko93,Wiltschko95,Wiltschko99,Maier,Zapka}. The underlying mechanism in such a RP compass is clearly of quantum mechanical nature, thus to what extent and under what conditions the quantum coherence or entanglement can play a positive role in RP compass has aroused great interest.

In the RP mechanism, due to the optical excitation the molecular conformation changes and the distance between two electrons increases. As a result, the electron-nuclear hyperfine interaction plays a dominant role instead of the exchange interaction. The singlet and triplet states are no longer the eigenstates of the RP Hamiltonian. Consequently, the singlet-triplet coherence is created and believed to be required for the RP navigation \cite{Walters,Kritsotakis,Imamoglu}. A quantitative connection between the compass sensitivity and the initial global electron-nuclear quantum coherence has been established, i.e., initial global coherence makes a more dominant contribution to the compass sensitivity as compared with local electronic coherence \cite{Cai4}. On the other hand, it has been pointed that the entanglement should last long enough to be used for bird's navigation \cite{Gauger,James1,James2}. And the interesting connections between the entanglement and the sensitivity of magnetic field intensity have also been found when the RP lifetime is not too long compared with the entanglement lifetime \cite{Cai1}. But for the singlet yield anisotropy (magnetic field direction sensitivity), quantum entanglement seems to have no direct contribution to it. The separable initial states can lead to more singlet yield anisotropy than the initial singlet state \cite{Cai1,Gauger}. Hore and his co-workers investigated the relation between compass properties and initial entanglement in detail, and found that it is somewhat complex \cite{Hannah}. For example, under certain conditions the initial entangled state can create the significant singlet yield anisotropy, but on the other condition the non-entangled initial states can lead to appreciable anisotropy \cite{Hannah}. Besides the roles of quantum coherence and entanglement, the effects of decoherence on the RP has also been investigated, and it has been found that some kinds of decoherence, can play positive roles in the RP compass \cite{Cai1,Cai2,Markus,Carrillo}, for example the performance of RP compass can be enhanced by the presence of correlated dephasing \cite{Cai2}. We can see that some conclusions above looks inconclusive, or even contradictive and what really determines the orientation, entanglement, coherence or someone else is still an open question.

In the RP mechanism, the hyperfine coupling which induces the singlet-triplet conversion depending on the magnetic field plays an essential role. The hyperfine interaction depends on the species of the nucleus and its location with respect to the electron wave function. The electron is usually influenced by the environment, and then the hyperfine coupling strength is not a constant but might fluctuate. Besides, there is ubiquitous external magnetic noise around the avian compass. So it is very important to investigate the effects of these noises on the RP navigation.

In this paper we investigate who determines the singlet yield, entanglement, coherence or someone else. We should note that quantifying coherence should be in a specific basis \cite{Baumgratz,Johan,Girolami}, and we find that the dark states of TEZE play a very important role in the RP compass. We define the quantum coherence in this dark state basis, and investigate its contribution to the singlet yield anisotropy. We prove that the dynamical process of the RP is an incoherent and local operation which can not create any coherence of the dark state of TEZE as well as any entanglement. Furthermore, we respectively calculate the singlet yields for two initial states: the coherent state (in the dark state basis) and the same state just removing the dark state coherence with the dark state population being preserved. For the later there are neither dark state coherence nor entanglement in the whole dynamical process. Surprisingly we find that in both cases the singlet yields are the same, and based on this result, we believe that the dark state population determines the singlet yield completely, and the dark state coherence and entanglement have little contribution to the singlet yield.

Also, we investigate the effects of hyperfine coupling noise and the magnetic noise on the singlet yield anisotropy. Although these noises are all inducing decoherence, their effects on the singlet yield anisotropy are significant different. The dark state population as well as the singlet yield anisotropy, is very fragile to the vertical magnetic noise, but is robust to and is even enhanced by the parallel magnetic noise. As for the hyperfine coupling noise, we find that the dark state population is very robust to the hyperfine noise, so that the orientation is very robust to the hyperfine noise.

\section{RP model and dark state}
\label{model}

The RP compass consists of two electronic spins coupled to an external magnetic field, and one of them interacts with the nuclei around it and the other is devoid of the hyperfine interaction. The hyperfine interaction provides asymmetry and leads to singlet-triplet transition required for the direction sensitivity. This model is verified by the RPs $[\mathbf{C}^{\bullet+}-\mathbf{P}-\mathbf{F}^{\bullet-}]$ \cite{Maeda} and $[\mathbf{FADH}^{\bullet}+\mathbf{O}^{\bullet-}_{\mathbf{2}}]$ \cite{Solov'yov}. The corresponding Hamiltonian is
\begin{equation}\label{H}
    \hat{H}_{0}=\gamma \textbf{B}\cdot(\hat{S}_{1}+\hat{S}_{2})+\sum_{n}\hat{S}_{1}\cdot \textbf{A}_{n}\cdot\hat{I}_{n},
\end{equation}
where $\hat{I}_{n}\equiv(I_{nx},I_{ny},I_{nz})$ is the nuclear spin operator, and $\textbf{A}_{n}$ is the anisotropic hyperfine tensor with a diagonal form $\textbf{A}_{n}=diag(A_{nx},A_{ny},A_{nz})$. And we consider an axially symmetric molecule, i.e., $A_{nx}=A_{ny}$. $\hat{S}_{i}\equiv(\sigma^{i}_{x},\sigma^{i}_{y},\sigma^{i}_{z})$ are the electronic spin operators ($i=1,2$), and $\gamma=\mu_{B}g_{s}/2$ is the gyromagnetic ratio, with $\mu_{B}$ being the Bohr magneton and $g_{s}=2$ being the $g$-factor of the electron. $\textbf{B}$ is the external magnetic field around the RP:
\begin{equation}\label{B}
\textbf{B}=B_{0}(\sin\theta \cos\phi, \sin\theta \sin\phi, \cos\theta),
\end{equation}
where $B_{0}$ is the intensity of the Earth's magnetic field, and $\theta$ and $\phi$ describe its orientation to the basis of the hyperfine interaction tensor. Due to the axial symmetry of the hyperfine tensor we set $\phi=0$ and focus on $\theta\in[0,\pi/2]$ without loss of generality. This is supported by the experiment that the avian compass does not depend on the polarity of magnetic field but only on its inclination \cite{Wiltschko72}. We consider the same singlet and triplet recombination rates, i.e., $k_{S}=k_{T}=k$, and in this case, the singlet yield can be calculated as
\begin{equation}\label{singlet yield}
    \Phi_{s}=\int_{0}^{\infty}r(t)f_{s}(t)dt,
\end{equation}
where $r(t)=k\exp(-kt)$ is the radical recombination probability distribution \cite{Steiner}, and $f_{s}(t)=\langle S|\rho_{s}(t)|S\rangle$ is the population of the singlet state $|S\rangle$. $\rho_{s}(t)=\mathrm{Tr}_{I}[U(t)\rho_{s}(0)\otimes \rho_{I}(0)U^{\dag}(t)]$ is the reduced electronic spin state at time $t$ with the partial trace over the nuclear subspace, where $U(t)=\exp[-i H_{0}t]$ is the evolution operator. It has been shown in different scenarios that $k$ should the order of $10^{4} \rm s^{-1}$, \cite{Gauger,Xu,Yang} so in this paper we let $k=10^{4} \rm s^{-1}$. The nuclei are initially in a completely mixed state, i.e., $\rho_{I}(0)=1/2^{N}\sum_{i}|i\rangle\langle i|$, and $N$ is the total number of the nuclei, and $|i\rangle$ is the basis of the nuclear environment. Generally we suppose that the electronic spins are initially in the singlet state $|S\rangle$ unless otherwise specified.

Through our calculation we find that the dark states (the corresponding eigenvalues are zero) of TEZE (the first term of Eq. (\ref{H})) play an important role in the RP model. Defining
\begin{equation}\label{M}
    M(\theta)=\sum_{i}\sin\theta\sigma^{i}_{x}+\cos\theta\sigma^{i}_{z},
\end{equation}
TEZE can be expressed as
\begin{equation}\label{HB}
    H_{B}=\gamma \textbf{B}\cdot(\hat{S}_{1}+\hat{S}_{2})=\gamma B_{0}M(\theta).
\end{equation}
The eigenvectors of $M(\theta)$ are as follows:
\begin{equation}\label{dark state}
\begin{split}
&|D^{1}(\theta)\rangle=|\psi_{1}(\theta)\rangle\otimes|\psi_{2}^{\perp}(\theta)\rangle,\\
     &|D^{2}(\theta)\rangle=|\psi_{1}^{\perp}(\theta)\rangle\otimes|\psi_{2}(\theta)\rangle, \\
    &|B^{1}(\theta)\rangle=|\psi_{1}(\theta)\rangle\otimes|\psi_{2}(\theta)\rangle,\\
    & |B^{2}(\theta)\rangle=|\psi^{\perp}_{1}(\theta)\rangle\otimes|\psi^{\perp}_{2}(\theta)\rangle
\end{split}
\end{equation}
with the eigenvalues 0, 0, 2, -2, respectively. $|\psi_{i}(\theta)\rangle=\cos\frac{\theta}{2}|1\rangle_{i}+\sin\frac{\theta}{2}|0\rangle_{i}$, $|\psi^{\perp}_{i}(\theta)\rangle=\sin\frac{\theta}{2}|1\rangle_{i}-\cos\frac{\theta}{2}|0\rangle_{i}$ are the eigenvectors of $\gamma\textbf{B}\cdot\hat{S}_{i}$ ($i=1,2$) with the corresponding eigenvalues $\gamma B_{0}$ and $-\gamma B_{0}$, respectively. $\sigma^{i}_{z}|1\rangle_{i}=|1\rangle_{i}$ and $\sigma^{i}_{z}|0\rangle_{i}=-|0\rangle_{i}$. Obviously, $|D^{1}(\theta)\rangle$ and $|D^{2}(\theta)\rangle$ are the dark states of $H_{B}$ and $M(\theta)$, and $|B^{1}(\theta)\rangle$ and $|B^{2}(\theta)\rangle$ are the bright states. From Eq. (\ref{dark state}) it can be seen that the dark states and the bright states are all product states, and have no any correlation between the two electrons. The singlet state $|S\rangle$ is invariant to rotations in the electron spin space, meaning that it is isotropic \cite{Hannah}. So that the singlet state can be expressed as
\begin{equation}\label{S}
\begin{split}
   |S\rangle&=\frac{1}{\sqrt{2}}\biggl(|\psi_{1}(\theta)\rangle\otimes|\psi^{\perp}_{2}(\theta)\rangle
   -|\psi^{\perp}_{1}(\theta)\rangle\otimes|\psi_{2}(\theta)\rangle\biggr) \\
            &=\frac{1}{\sqrt{2}}\biggl(|D^{1}(\theta)\rangle-|D^{2}(\theta)\rangle\biggr)
\end{split}
\end{equation}
which only depends on the dark states. Thus the singlet state population can be divided into two parts, i.e.,
\begin{equation}\label{fs}
    f_{s}(t)=f_{p}(t)+f_{c}(t),
\end{equation}
where
\begin{equation}\label{fsd}
    f_{p}(t)=\frac{1}{2}\biggl(\langle D^{1}(\theta)|\rho_{s}(t)|D^{1}(\theta)\rangle+\langle D^{2}(\theta)|\rho_{s}(t)|D^{2}(\theta)\rangle\biggr)
\end{equation}
and
 \begin{equation}\label{fsc}
 \begin{split}
    f_{c}(t)=-\frac{1}{2}\biggl(\langle D^{1}(\theta)|\rho_{s}(t)|D^{2}(\theta)\rangle
    +\langle D^{2}(\theta)|\rho_{s}(t)|D^{1}(\theta)\rangle\biggr).
 \end{split}
\end{equation}
We define $f_{c}(t)$ and $f_{p}(t)$ as the dark state coherence and population at time $t$, respectively, whose contributions to the singlet yield are:
\begin{equation}\label{php}
  \Phi_{p}=\int_{0}^{\infty}r(t)f_{p}(t)dt,
\end{equation}
\begin{equation}\label{phc}
  \Phi_{c}=\int_{0}^{\infty}r(t)f_{c}(t)dt.
\end{equation}
Obviously, $\Phi_{s}=\Phi_{p}+\Phi_{c}$. The essential of the orientation is the singlet yield anisotropy, i.e., the singlet yield $\Phi_{s}$ is different for different $\theta$.
\section{Contributions of the dark state coherence and population to the singlet yield}
\label{contribution}
We firstly consider a simple case that only the vertical hyperfine coupling is considered, i.e., $A_{nx}=A_{ny}=0$. The nuclear spins can then be treated as inducing an effective magnetic field (depending on their initial states) for the electron spin. Although this model is very simple, the basic physical process for the magnetoreception holds. And such a simple model allows us to obtain analytic results which are quite useful for understanding the essential effects of the dark state coherence and population.

Now we consider the most basic RP model that there is only one nucleus around the electron, i.e., $N=1$ (the multi-nuclei RP model is discussed in Appendix A). If the nuclear spin is in the up (down) state $|\uparrow\rangle$ ($|\downarrow\rangle$), the effective magnetic field is $A_{z}\hat{\textbf{z}}/\gamma$ ($-A_{z}\hat{\textbf{z}}/\gamma$) with $\hat{\textbf{z}}$ being the $z$ direction. The dark state population and coherence at time $t$ can be calculated as
\begin{eqnarray}\label{fp}
     f_{p}(t)=\frac{1}{2}\biggl(f_{p+}(t)+f_{p-}(t)\biggr), \\
     f_{c}(t)=\frac{1}{2}\biggl(f_{c+}(t)+f_{c-}(t)\biggr).
\end{eqnarray}
where
\begin{equation}\label{fp1}
    f_{p\pm}(t)=\frac{1}{2}\biggl[1-\frac{1}{2}\sin^{2}(\theta_{\pm}-\theta)\bigl[1-\cos (2\omega_{\pm}t)\bigr]\biggr]
\end{equation}
and
\begin{equation}\label{fc1}
\begin{split}
    f_{c\pm}(t)&=\frac{1}{2}\biggl[\cos^{4}\frac{\theta_{\pm}-\theta}{2}\cos\bigl[2(\omega_{\pm}-\omega_{0})t\bigr] \\
                 & +\sin^{4}\frac{\theta_{\pm}-\theta}{2}\cos\bigl[2(\omega_{\pm}+\omega_{0})t\bigr] \\
                 &+\frac{1}{2}\sin^{2}(\theta_{\pm}-\theta)\cos(2\omega_{0}t)\biggr].
\end{split}
\end{equation}
The symbol $+$ ($-$) means that the initial nuclear spin state is $|\uparrow\rangle$ ($|\downarrow\rangle$). $B_{\pm}=\sqrt{B^{2}_{x}+(B_{z}\pm A_{z}/\gamma)^{2}}$, $\omega_{\pm}=\gamma B_{\pm}$, $\sin\theta_{\pm}=B_{x}/B_{\pm}$, $\cos\theta_{\pm}=(B_{z}\pm A_{z}/\gamma)/B_{\pm}$, $\omega_{0}=\gamma B_{0}$. From Eq. (\ref{fp1}) (Eq. (\ref{fc1})) we can see that $f_{p}(t)$ ($f_{c}(t)$) oscillates with $2\omega_{\pm}$ ($2(\omega_{\pm}-\omega_{0})$, $2(\omega_{\pm}+\omega_{0})$ and $2\omega_{0}$). It has been shown that the hyperfine coupling strength should be stronger than the geomagnetic field intensity \cite{Xu}. When the hyperfine coupling is relatively strong compared with the geomagnetic field, $\omega_{\pm}-\omega_{0}$, $\omega_{\pm}+\omega_{0}$, $\omega_{\pm}$ and $\omega_{0}\gg k$. The time integral of these high-frequency oscillation terms of $f_{c}(t)$ and $f_{p}(t)$ approximately equal to zero, i.e., they have little contribution to the singlet yield. Neglecting the high-frequency oscillations, $f_{p}(t)$ and $f_{c}(t)$ can be expressed as
\begin{equation}\label{fp2}
    f_{p}(t)\approx\frac{1}{2}-\frac{1}{4}\sin^{2}(\theta_{\pm}-\theta)
\end{equation}
and
\begin{equation}\label{fc2}
\begin{split}
    f_{c}(t)\approx0.
\end{split}
\end{equation}
Substituting Eq. (\ref{fp2}) (Eq. (\ref{fc2})) into Eq. (\ref{php}) (Eq. (\ref{phc})), we obtain
\begin{equation}\label{phip2}
  \Phi_{p}\approx\Phi_{s}\approx\frac{1}{2}-\frac{1}{4}\sin^{2}(\theta_{\pm}-\theta),
\end{equation}
\begin{equation}\label{phic2}
  \Phi_{c}\approx0.
\end{equation}
$\Phi_{c}$ is always zero and $\Phi_{p}$ is always equal to $\Phi_{s}$, which means that the singlet yield is determined completely by the dark state population and is independent of the dark state coherence.

\begin{figure}[tbp]
\centering \includegraphics[width=8cm]{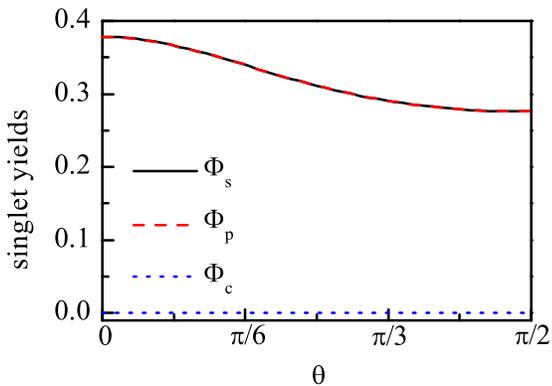}
\hspace{0.5cm} \caption{(Color online) The singlet yields $\Phi_{s}$ (black-solid line), $\Phi_{c}$ (blue-dotted line) and $\Phi_{p}$ (red-dashed line) as functions of $\theta$ for $B_{0}=46\mu\rm T$, $A_{z}=5\Lambda$ and $A_{x}=A_{y}=3\Lambda$. It should be noted that $\Phi_{p}$ coincides with $\Phi_{s}$.}
\end{figure}

Now we consider the horizontal hyperfine interaction, i.e., $A_{nx}=A_{ny}\neq0$ and in this case the analytical result can not be obtained. For convenience of our discussion, we define  $\Lambda\equiv\gamma\times46\mu \rm T$ as the energy scale, which is the electronic spin energy induced by the geomagnetic field of $46\mu \rm T$ in Frankfurt \cite{Ritz09}. For the single-nucleus RP model ($N=1$) we consider $A_{x}=A_{y}=3\Lambda$, $A_{z}=5\Lambda$ and $B_{0}=46\mu \rm T$ as an example and numerically calculate the singlet yield as shown in Fig. 1. From Fig. 1 we can see that $\Phi_{c}$ is always zero and $\Phi_{p}$ always coincides with $\Phi_{s}$ along different directions, which means that the dark state population determines the singlet yield completely. In Appendix B we consider two important experiments that the bird can adapt to different field intensities \cite{Wiltschko72,Wiltschko78,Wiltschko06,Wiltschko13} and the weak oscillating field can completely disorient the bird \cite{Ritz04,Ritz05,Ritz09,Wiltschko15} which support the RP mechanism, and we also explain these experimental results from the point view of the dark state population.

Recently, Hore and his coworkers \cite{Hannah} have pointed out that the singlet yield anisotropy (the singlet yield is different for different $\theta$), is essential to the magnetic sensitivity, which not only can come from the anisotropic hyperfine interaction but also can come from the anisotropic initial state. More specifically, if the initial state is isotropic (for example the singlet state $|S\rangle$), the anisotropic hyperfine interaction ($A_{nx}=A_{ny}\neq A_{nz}$) can induce the singlet yield anisotropy; and if the hyperfine interaction is isotropic ($A_{nx}=A_{ny}=A_{nz}$), the anisotropic initial state (for example the triplet state $|T_{0}\rangle=(|10\rangle+|01\rangle)/\sqrt{2}$) can also induce the yield anisotropy. In Appendix C we also consider the singlet yield anisotropy coming from the anisotropic initial state, and the same conclusion that the singlet yield is completely determined by the dark state population is also arrived at.

Although the singlet yield is completely determined by the dark state population, we can not yet draw a conclusion that the dark state coherence has no contribution to it, because there is the dark state coherence in the initial state (the singlet yield $|S\rangle$) and we do not know whether the initial dark state coherence influences the dark state population in the dynamics, so that influences the singlet yield indirectly. To answer this question, we can remove the initial dark state coherence, i.e., using an incoherent state. But only removing the initial dark state coherence is not enough, since it is not sure whether the dynamical process is an incoherent operation or not (i.e., whether the dynamical process generate the dark state coherence from an incoherent state or not). A completely positive trace preserving map $\Lambda$ is said to be an incoherent operation if it can be written as $\Lambda (\rho)=\sum_{l}K_{l}\rho K^{\dag}_{l}$ with the incoherent Kraus operators mapping every incoherent state to some other incoherent states, i.e., $K_{l}\mathcal{I} K^{\dag}_{l}\subseteq \mathcal{I}$, where $\mathcal{I}$ is the set of incoherent states and $\sum_{l}K^{\dag}_{l} K_{l}=\mathbb{I}$ \cite{Baumgratz}.

In the RP model, the dynamical map can be expressed as
\begin{equation}\label{map}
 \Lambda \bigl(\rho_{s}(0)\bigr)=\sum_{ij}K_{ij}\rho_{s}(0)(K_{ij})^{\dag},
\end{equation}
with the Krause operator being
\begin{equation}\label{krause}
 K_{ij}=\frac{1}{2^{N/2}}U^{ij}_{1}(t) U_{2}(t),
\end{equation}
where $U^{ij}_{1}(t)=\langle i|U_{1}(t)|j\rangle$, $U_{1}(t)=\exp[-i(\gamma \textbf{B}\cdot\hat{S}_{1}+\sum_{n}\hat{S}_{1}\cdot \textbf{A}_{n}\cdot\hat{I}_{n})t]$, $U_{2}(t)=\exp[-i\gamma \textbf{B}\cdot\hat{S}_{2}t]$, and $|i\rangle$ (or $|j\rangle$) is the basis of the nuclear bath. In the dark state basis, the incoherent state can be expressed as $\rho_{in}=p_{1}|D^{1}\rangle\langle D^{1}|+p_{2}|D^{2}\rangle\langle D^{2}|$ with $p_{1}+p_{2}=1$. It can be proved that
\begin{equation}\label{in-krause}
  \mathrm{Tr}\bigl[K_{ij}\rho_{in}(K_{ij})^{\dag}|D^{m}\rangle\langle D^{m'}|\bigr]=0,
\end{equation}
where $m,~m'=1,2$ and $m\neq m'$. The Krause operator $K_{ij}$ can not produce the dark state coherence from the incoherent states, or the dynamical dark state coherence completely comes from the initial dark state coherence. In this sense, the dynamical map (Eq. (\ref{map})) is an incoherent operation.

If we remove the dark state coherence from the initial singlet state $|S\rangle$, i.e., consider the incoherent state $\rho^{in}_{s}(0)=\frac{1}{2}|D^{1}\rangle\langle D^{1}|+\frac{1}{2}|D^{2}\rangle\langle D^{2}|$, according to the discussion above the dark state coherence in the dynamics is obviously zero and the dark state population at time $t$ is
\begin{equation}\label{fsp_in}
\begin{split}
f^{in}_{p}(t)&=\frac{1}{2}\biggl(\langle D^{1}(\theta)|\Lambda \bigl(\rho^{in}_{s}(0)\bigr)|D^{1}(\theta)\rangle\\
    &+\langle D^{2}(\theta)|\Lambda \bigl(\rho^{in}_{s}(0)\bigr)|D^{2}(\theta)\rangle\biggr)\\
    &=f_{p}(t).
\end{split}
\end{equation}
Interestingly, the dark state population are not influenced and its contribution to the singlet yield is
\begin{equation}\label{phip_in}
  \Phi^{in}_{p}=\Phi_{p}\approx\Phi_{s}.
\end{equation}
It can be seen that whether removing the initial dark state coherence or not, the singlet yield comes from the dark state population can not be influenced and is equal to $\Phi_{s}$. This means that the initial dark state coherence have little contribution to the singlet yield.

We also note that because the dark states are all the product states for two electrons, the incoherent state $\rho_{in}=p_{1}|D^{1}\rangle\langle D^{1}|+p_{2}|D^{2}\rangle\langle D^{2}|=p_{1}|\psi_{1}\rangle\langle \psi_{1}|\otimes|\psi^{\perp}_{2}\rangle\langle\psi^{\perp}_{2}|+p_{2}|\psi^{\perp}_{1}\rangle\langle \psi^{\perp}_{1}|\otimes|\psi_{2}\rangle\langle\psi_{2}|$ is a separable state, and has no any quantum correlation (entanglement). And the dynamical map (Eq. (\ref{map})) is a local operation due to no interaction between the two electrons, and only map the separable state (incoherent state) $\rho_{in}$ to another separable state (incoherent state)
\begin{equation}\label{}
\begin{split}
  \Lambda \bigl(\rho_{in}\bigr)&=\frac{p_{1}}{2^{N}}\sum_{ij}U^{ij}_{1}(t)|\psi_{1}\rangle\langle \psi_{1}|(U^{ij}_{1}(t))^{\dag}\otimes|\psi^{\perp}_{2}\rangle\langle\psi^{\perp}_{2}|  \\
  &+\frac{p_{2}}{2^{N}}\sum_{ij}U^{ij}_{1}(t)|\psi^{\perp}_{1}\rangle\langle \psi^{\perp}_{1}|(U^{ij}_{1}(t))^{\dag}\otimes|\psi_{2}\rangle\langle\psi_{2}|.
\end{split}
\end{equation}
So it can be concluded that the dynamical map $\Lambda$ is an incoherent and local operation that can not create any dark state coherence as well as any quantum correlation (entanglement). From Eqs. (\ref{fsp_in}) and (\ref{phip_in}), it can be seen that if we consider an initial separable state (incoherent state) $\rho^{in}_{s}(0)=\frac{1}{2}|D^{1}\rangle\langle D^{1}|+\frac{1}{2}|D^{2}\rangle\langle D^{2}|$ compared with the initial singlet state $|S\rangle$ (an entangled state), the dark state population as well as the singlet yield remains the same. Thus another interesting result, quantum correlations have little contribution to the magnetic sensitivity, is obtained. So it can be concluded that the dark state population makes the main contribution to the magnetic sensitivity, and the dark state coherence and entanglement have little contribution to it. It should be noted that it is only the dark state coherence has little contribution to the time integrated singlet yield, but if we consider quantum coherence in other basis, it may play a certain role. For example, the singlet-triplet coherence is believed to be required for the RP navigation \cite{Baumgratz,Johan,Girolami}.

One should note that according to Eq. (\ref{fsc}) the dark state cohernce $f_{c}(t)$ can be expressed as
\begin{equation}\label{}
\begin{split}
 & f_{c}(t)=\\
 & -\frac{1}{2^{N}}\mathrm{Re}\biggl[e^{i2\omega_{0}t}\langle D^{1}(\theta)|\sum_{ij}U^{ij}_{1}(t)\rho_{s}(0)(U^{ij}_{1}(t))^{\dag}|D^{2}(\theta)\rangle\biggr].
\end{split}
\end{equation}
For any initial state $\rho_{s}(0)$, and any interaction $\sum_{n}\hat{S}_{1}\cdot \textbf{A}_{n}\cdot\hat{I}_{n}$ (i.e, any time evolution operator $U_{1}(t)$) between electron 1 and the corresponding nuclear bath, the dark state coherence $f_{c}(t)$ has a fixed oscillating factor $e^{i2\omega_{0}t}$ where the frequency $2\omega_{0}$ is far greater than $k$, so that it does not contribute to the singlet yield. This means that in the present RP model, the conclusion that it is the dark state population rather than the dark state coherence and entanglement determines the singlet yield, is independent of the hyperfine interaction and the initial RP state.

That the singlet yield is completely determined by the dark state population and has nothing to do with the dark state coherence and entanglement can be understood as follows. Generally, the nuclear environment limits the time scale of coherence behavior to $\tau\sim 1/\bar{A}=N/\sum|\textbf{A}_{n}|\sim10^{-8}s$ which is much shorter than the RP lifetime $1/k=10^{-4}\rm s$, so that the dark state coherence has no time to contribute to the magnetic sensitivity, which is different from the magnetometry based on diamond in which the electronic coherence plays an essential role in the magnetic sensitivity \cite{Taylor,Maze,Balasubramanian}. In another word, if the RP lifetime is approximately equal to or shorter than $\tau$, i.e., the recombination rate $k$ is sufficiently large, the dark state coherence even quantum correlation can have enough time to contribute to the singlet yield. The connection between the entanglement and the magnetic sensitivity is established when the RP lifetime is supposed to be not too long, such as $k=5.8\times10^{8}\rm s^{-1}$, compared with the entanglement lifetime \cite{Cai1}. In this paper if we also set $k=5.8\times10^{8}\rm s^{-1}$, the dark state coherence will contribute to the singlet yield, because the lifetime scale $1/k=(1/5.8)\times10^{-8}\rm s$ is the same order of $\tau\sim 1/\bar{A}$ for $\bar{A}\sim5\Lambda$, and the dark state coherence and entanglement can have enough time to contribute to the singlet yield.

\section{Effects of the noises}
There are ubiquitous noises around the RP, and the investigation of the effects of the noises on the magnetic sensitivity has both theoretical and practical significance. The Hamiltonian of the noise can be expressed as
\begin{equation}
  \hat{H}'(t)=h(t)\hat{h}.
\end{equation}
Considering a Gaussian white noise, i.e., $\langle h(t)\rangle=0$ and $\langle h(t)h(\tau)\rangle=\Gamma\delta(t-\tau)$, and after some derivations (see Appendix D) we can obtain the standard master equation
\begin{equation}\label{drhot}
\frac{d}{dt}\rho(t)=-i[H_{0},~\rho(t)]-\Gamma[\hat{h},~[\hat{h},~\rho(t)]].
\end{equation}

\subsection{magnetic noise}

\begin{figure}[tbp]
\centering \includegraphics[width=8.0cm]{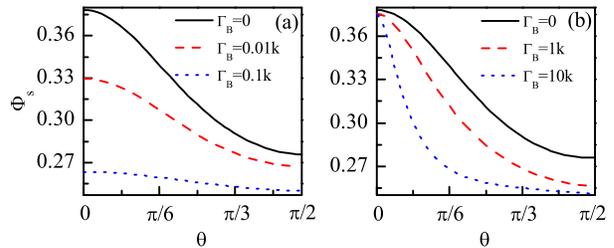}
\hspace{0.5cm} \caption{(Color online) The singlet yield $\Phi_{s}$ as functions of the direction angle $\theta$ for (a) the vertical magnetic noises and (b) the parallel magnetic noises. $A_{z}=5\Lambda$, $A_{x}=A_{y}=3\Lambda$, $B_{0}=46\rm\mu T$.}
\end{figure}

Firstly we consider the magnetic noise due to its ubiquity around the world. The magnetic noise Hamiltonian can be expressed as
\begin{equation}\label{magnetic noise}
     H'(t)=\gamma\textbf{B}'(t)\cdot(\hat{S}_{1}+\hat{S}_{2}),
\end{equation}
with
\begin{equation}
     \textbf{B}'(t)=B'(t)(\sin\vartheta \cos\varphi, \sin\vartheta \sin\varphi, \cos\vartheta)
\end{equation}
being the fluctuating field, where $B'(t)$ describes the strength of the fluctuating field and $\vartheta$ and $\varphi$ are its direction angles. We also set $\varphi=0$ due to the axial symmetry of the hyperfine interaction tensor. In this case the magnetic noise Hamiltonian can be expressed as $H'(t)=\gamma B'(t)M(\vartheta)$. The form of $M(\vartheta)$ is given in Eq. (\ref{M}), and the only difference is replacing $\theta$ by $\vartheta$. Here we consider a Gaussian white noise, i.e., $\langle \gamma B'(t)\rangle=0$ and $\langle \gamma B'(t)\gamma B'(\tau)\rangle=\Gamma_{B}\delta(t-\tau)$. According to Eq. (\ref{drhot}) we can obtain the master equation:
\begin{equation}
\frac{d}{dt}\rho(t)=-i[H_{0},~\rho(t)]-\Gamma_{B}[M(\vartheta),~[M(\vartheta),~\rho(t)]].
\end{equation}

Here, two kinds of fields are investigated: the parallel fluctuating field, i.e., $\vartheta=\theta$, and the vertical fluctuating field, i.e., $\vartheta=\theta+\pi/2$ ($\theta$ is the direction of the geomagnetic field). For simplicity, we only consider the single-nucleus RP model, i.e., $N=1$. Considering $B_{0}=46\rm\mu T$, $A_{z}=5\Lambda$ and $A_{x}=A_{y}=3\Lambda$ as an example, we numerically calculate $\Phi_{s}$ for the vertical ($\vartheta=\theta+\pi/2$) and parallel ($\vartheta=\theta$) magnetic noises, and show the results in Fig. 2. From Fig. 2(a) we can see that if the vertical magnetic noise is approximately equal to or larger than $0.1k$, the singlet yield profile flattens out and thus the magnetic sensitivity is destroyed completely. Because $M(\vartheta)|D^{1}(\vartheta)\rangle=0$, $M(\vartheta)|D^{2}(\vartheta)\rangle=0$, $M(\vartheta)|B^{1}(\vartheta)\rangle=2|B^{1}(\vartheta)\rangle$ and $M(\vartheta)|B^{2}(\vartheta)\rangle=-2|B^{2}(\vartheta)\rangle$, the noise only decays $\langle B^{1}(\vartheta)|\rho(t)|B^{2}(\vartheta)\rangle$, $\langle B^{2}(\vartheta)|\rho(t)|B^{1}(\vartheta)\rangle$, $\langle B^{1}(\vartheta)|\rho(t)|B^{1}(\vartheta)\rangle$ and $\langle B^{2}(\vartheta)|\rho(t)|B^{2}(\vartheta)\rangle$. If the magnetic noise is vertical to the geomagnetic field, i.e., $\vartheta=\theta+\pi/2$, $|B^{i}(\vartheta)\rangle$ overlaps with $|D^{i'}(\theta)\rangle$ the dark state of $H_{B}$ (see Eq. (\ref{dark state})) ($i,i'=1,2$). So a part of dark state population will decay with the decaying matrix elements in the basises $|B^{1}(\vartheta)\rangle$ and $|B^{2}(\vartheta)\rangle$. As a result although the RP exists (the RP lifetime is $1/k$) there is no or no enough dark state population to create the singlet yield. So the vertical magnetic noise should be weak enough (for example $\Gamma_{B}=0.01k$) in this way there is enough dark state population to induce the singlet yield.

From Fig. 2(b), we can see that the magnetic sensitivity is more robust to the parallel magnetic noise than to the vertical magnetic noise. Only the parallel noise is approximately equal to or larger than $10k$ (which is much larger than $0.1k$ for which the vertical magnetic noise destroys the magnetic sensitivity significantly), can the singlet yield flattens out for large angles. Interestingly, the magnetic sensitivity can be enhanced by the parallel magnetic noise (for example $\Gamma_{B}=1k$). This can be understood as follows: When $\vartheta=\theta$, the eigenvectors of $M(\vartheta)$ is the same as those of $H_{B}$ (see Eq. (\ref{HB})), and interestingly the dark states of $H_{B}$ expand a subspace which is immune to the parallel magnetic noise. Thus the dark state population is more robust to the parallel magnetic noise. However, due to the hyperfine interaction the dark states can be transferred into the bright states (specifically the dark state $|D^{i}(\theta)\rangle$ is transferred into the bright state $|B^{i}(\theta)\rangle$, $i=1,2$) which will be decayed by the parallel magnetic noise. So if the parallel magnetic noise is too strong the dark state population will be decreased and then the magnetic sensitivity will be disrupted.

\subsection{hyperfine coupling noise}
It is well known that the hyperfine interaction is essential to the RP compass, and is related to the electron envelope function. The electron can be influenced by the inevitable environment, thus the hyperfine coupling strength is not a constant but can fluctuate. We define the hyperfine coupling noise as
\begin{equation}\label{hyperfine noise}
    H'(t)=\hat{I}\cdot\textbf{A}'(t)\cdot\hat{S}_{1}.
\end{equation}
Here, we only consider the single-nucleus RP model and the fluctuations for different directions being the same, i.e., $A'_{x}(t)=A'_{y}(t)=A'_{z}(t)=A'(t)$. In this case the hyperfine coupling noise Hamiltonian can be expressed as $H'(t)=A'(t)\hat{I}\cdot\hat{S}_{1}$. Here we also consider the Gaussian white noise, i.e., $\langle A'(t)\rangle=0$ and $\langle A'(t)A'(\tau)\rangle=\Gamma_{H}\delta(t-\tau)$.  According to Eq. (\ref{drhot}) we can obtain the master equation:
\begin{equation}
\frac{d}{dt}\rho(t)=-i[H_{0},~\rho(t)]-\Gamma_{H}[\hat{I}\cdot\hat{S}_{1},~[\hat{I}\cdot\hat{S}_{1},~\rho(t)]].
\end{equation}

Considering $B_{0}=46\rm\mu T$, $A_{z}=5\Lambda$ and $A_{x}=A_{y}=3\Lambda$, we calculate the singlet yields for different strengths of the hyperfine coupling noise, and plot Fig. 3 to show the results. From Fig. 3 we can see that the hyperfine coupling noise for $\Gamma_{H}=1k$ almost does not influence the singlet yield, and even for $\Gamma_{H}=10k$ the magnetic sensitivity is still very robust. The RP compass is more robust to the hyperfine coupling noise compared with the magnetic noise. So we can conclude that although the electron can be influenced by the inevitable environment and then the hyperfine interaction is influenced, the RP compass can still orient. This can be understand from the point view of the dark state population, i.e., the dark state population is very robust to the hyperfine noise, so that the orientation is very robust to the hyperfine noise.

\begin{figure}[tbp]
\centering \includegraphics[width=8.0cm]{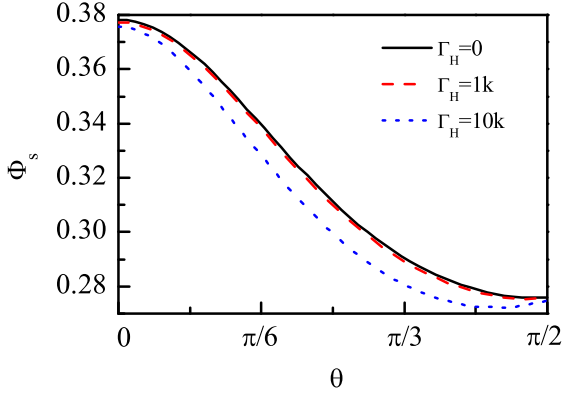}
\hspace{0.5cm} \caption{ (Color online) The singlet yield $\Phi_{s}$ as functions of the direction angle $\theta$ for the hyperfine coupling noises. $A_{z}=5\Lambda$, $A_{x}=A_{y}=3\Lambda$, $B_{0}=46\rm\mu T$.}
\end{figure}

\section{Discussions and Conclusions}
\label{conclusions}

In this paper we have investigated who among quantum entanglement, coherence or someone else, determines the singlet yield. We have found that the dark states of TEZE play a very important role in the singlet yield. In this dark state basis, we have proved that the dynamical process is an incoherent and local operation that can not produce any dark state coherence as well as any entanglement. Then we have calculated the singlet yields for two initial states: the coherent state (in the dark state basis) and the same state just removing the dark state coherence where the dark state population is preserved. For the later there are neither dark state coherence nor entanglement in the whole dynamical process. Surprisingly we have found that in both cases the singlet yields are the same, and based on these results, it can be concluded that the dark state population determines the singlet yield completely, and the dark state coherence and entanglement have little contribution to the singlet yield. The dark state coherence and entanglement have little contribution to the singlet yield can be understood as follows: In the present RP magnetoreception model, the nuclei around the electron limit the time scale of the coherence behavior (or the entanglement) to $\tau\sim 1/\bar{A}=N/\sum|\textbf{A}_{n}|$ ($\tau\sim10^{-8}\rm s$ for $\bar{A}\sim \gamma B_{0}$) which is much shorter than the RP lifetime $1/k=10^{-4}\rm s$, so that the dark state coherence have no enough time to contribute to the singlet yield. Due to the spin relaxation, some real RPs can not sustain for $10^{-4}\rm s$ but generally for $10^{-6}s$ which is also far greater than $\tau\sim10^{-8}s$. Thus if we set $k=10^{6}s^{-1}$, our results above are still valid. Finally, we have investigated the effects of the magnetic field and the hyperfine coupling noises. The vertical magnetic noise decreases the dark state population significantly and then disrupts the singlet yield anisotropy dramatically. However the singlet yield anisotropy is robust to the parallel magnetic noise and can be even enhanced, because the dark states expand a subspace which is immune to the parallel noise. And the magnetic sensitivity is more robust to the hyperfine coupling noise, so that although the electron can be influenced by the inevitable environment and then the hyperfine interaction essential to the magnetic sensitivity is influenced, the RP compass can still orient.

\section{Acknowledgements}

This work was supported by the National Natural Science Foundation of China (Grant No. 11274043).

\appendix
\section{Multi-nuclei RP model}
In the main text, the single-nucleus RP model has been discussed, and now we consider the multi-nuclei RP model. Firstly, we only consider the vertical hyperfine interaction, i.e., $A_{nx}=A_{ny}=0$ for all $n$. For simplicity, we assume that the hyperfine interaction strengths for all the nuclei are the same, i.e., $A_{nz}=T_{z}/2$ for all $n$. In this case the Hamiltonian (see Eq. (1)) in the main text can be expressed as $\hat{H}_{0}=\gamma \textbf{B}\cdot(\hat{S}_{1}+\hat{S}_{2})+T_{z}\hat{S}_{1z}\hat{J}_{z}$ with $\hat{J}_{z}=\sum_{n}\hat{I}_{nz}/2$. And the initial nuclear state (the completely mixed state) can be expressed as $\rho_{I}(0)=\frac{1}{2^{N}}\sum_{J}\sum_{M}\nu(N,J)|J,M\rangle\langle J,M|$ where $\nu(N,J)=\binom{N}{N/2-J}-\binom{N}{N/2-J-1}$ denotes the degeneracy of the spin bath with $\binom{N}{-1}=0$ \cite{Schliemann}. $|J,M\rangle$ is the eigenvector of $\hat{J}_{z}$ with $\hat{J}_{z}|J,M\rangle=M|J,M\rangle$, where $J=0,1,2,\cdot\cdot\cdot,N/2$ for $N$ being even, $J=1/2,3/2,\cdot\cdot\cdot,N/2$ for $N$ being odd ($N$ is the total number of the nuclei), $M=-J,-J+1,\cdot\cdot\cdot,J-1,J$ \cite{Gross}. In this case the effective field induced by the nuclear spins is $MT_{z}\hat{\textbf{z}}/\gamma$. The dark state coherence and population can be calculated as
\begin{equation}\label{fdm}
\begin{split}
    f_{p}(t)=\frac{1}{2^{N}}\sum_{J}\sum_{M}\nu(N,J)f^{p}_{M}(t),\\
    f_{c}(t)=\frac{1}{2^{N}}\sum_{J}\sum_{M}\nu(N,J)f^{c}_{M}(t)
\end{split}
\end{equation}
with
\begin{equation}
    f^{p}_{M}(t)=\frac{1}{2}-\frac{1}{4}\sin^{2}(\theta_{M}-\theta)\bigl[1-\cos (2\omega_{M}t)\bigr]
\end{equation}
and
\begin{equation}\label{fcmm}
\begin{split}
    f^{c}_{M}(t)&=\frac{1}{2}\cos^{4}\frac{\theta_{M}-\theta}{2}\cos\bigl[2(\omega_{M}-\omega_{0})t\bigr] \\
               &+\frac{1}{2}\sin^{4}\frac{\theta_{M}-\theta}{2}\cos\bigl[2(\omega_{M}+\omega_{0})t\bigr] \\
               &+\frac{1}{4}\sin^{2}(\theta_{M}-\theta)\cos(2\omega_{0}t),
\end{split}
\end{equation}
where $\omega_{M}=\gamma B_{M}$, $B_{M}=\sqrt{B^{2}_{x}+(B_{z}+MT_{z}/\gamma)^{2}}$, $\sin\theta_{M}=B_{x}/B_{M}$, $\cos\theta_{M}=(B_{z}+MT_{z}/\gamma)/B_{M}$.

For odd nuclear number $N$, ignoring the hight frequency oscillating terms, $f_{p}(t)$ and $f_{c}(t)$ can be expressed as
\begin{equation}\label{fpM_odd}
    f_{p}(t)\approx\frac{1}{2^{N+1}}\sum_{J}\sum_{M}\nu(N,J)\biggl[1-\frac{1}{2}\sin^{2}(\theta_{M}-\theta)\biggr]
\end{equation}
and
\begin{equation}\label{fcM_odd}
    f_{c}(t)\approx0.
\end{equation}
Substituting Eq. (\ref{fpM_odd}) (Eq. (\ref{fcM_odd})) into Eq. (11) (Eq. (12)) in the main text, we obtain
\begin{equation}\label{AMz}
\begin{split}
    &\Phi_{c}\approx0, \\
    &\Phi_{p}\approx\Phi_{s}\approx\frac{1}{2^{N+1}}\sum_{J}\sum_{M}\nu(N,J)\biggl[1-\frac{1}{2}\sin^{2}(\theta_{M}-\theta)\biggr].
\end{split}
\end{equation}
$\Phi_{c}$ is always zero and $\Phi_{p}$ is always equal to $\Phi_{s}$, which means that the singlet yield is determined completely by the dark state population.

For even nuclear number $N$, $M$ can equal to 0, so that the hyperfine interaction has no effect on the electrons, i.e., the effective field $MT_{z}\hat{\textbf{z}}/\gamma$ induced by the nuclear spins is zero. In this case, $\omega_{M=0}=\omega_{0}$ and $\theta_{M=0}=\theta$. And there exists a constant term $1/2$ in $f^{c}_{M}(t)$ (see Eq. (\ref{fcmm})). Neglecting the hight frequency oscillating terms, $f_{p}(t)$ and $f_{c}(t)$ can be expressed as
\begin{equation}\label{fpM_even}
    f_{p}(t)\approx\frac{1}{2}-\frac{1}{2^{N+2}}\sum_{J}\sum_{M}\nu(N,J)\sin^{2}(\theta_{M}-\theta)
\end{equation}
and
\begin{equation}\label{fcM_even}
    f_{c}(t)\approx\frac{1}{2^{N+1}}\frac{N!}{((N/2)!)^{2}}.
\end{equation}
Substituting Eq. (\ref{fpM_even}) (Eq. (\ref{fcM_even})) into Eq. (11) (Eq. (12)) in the main text, we obtain
\begin{equation}\label{AMz1}
\begin{split}
    &\Phi_{c}\approx\frac{1}{2^{N+1}}\frac{N!}{((N/2)!)^{2}},\\
    &\Phi_{p}\approx\frac{1}{2}-\frac{1}{2^{N+2}}\sum_{J}\sum_{M}\nu(N,J)\sin^{2}(\theta_{M}-\theta).
\end{split}
\end{equation}
$\Phi_{c}$ is always a constant and is independent of the geomagnetic field. As a result, although the dark state coherence has contribution to the singlet yield, it does not contribute to the magnetic sensitivity ($\partial\Phi_{c}/\partial\theta=0$). This can be understood as follows. When $M=0$, two electrons are only influenced by the geomagnetic field, and there is no hyperfine interaction to induce the transition between the singlet and triplet states. Therefore for the initial singlet state, its population is unchanged, and thus $\Phi_{c}$ is a constant ($\frac{1}{2^{N+1}}\frac{N!}{((N/2)!)^{2}}$). Although $\Phi_{p}$ is not equal to $\Phi_{s}$, $\partial\Phi^{0}_{p}/\partial\theta=\partial\Phi_{s}/\partial\theta$, in another word the magnetic sensitivity is completely determined by $f_{p}(t)$. It is noted that $\Phi_{c}$ is decreasing with the increasing $N$, and for a sufficiently large $N$ the singlet yield which comes from the dark state coherence will disappear.

\begin{widetext}
\begin{center}
\includegraphics[width=12.0cm]{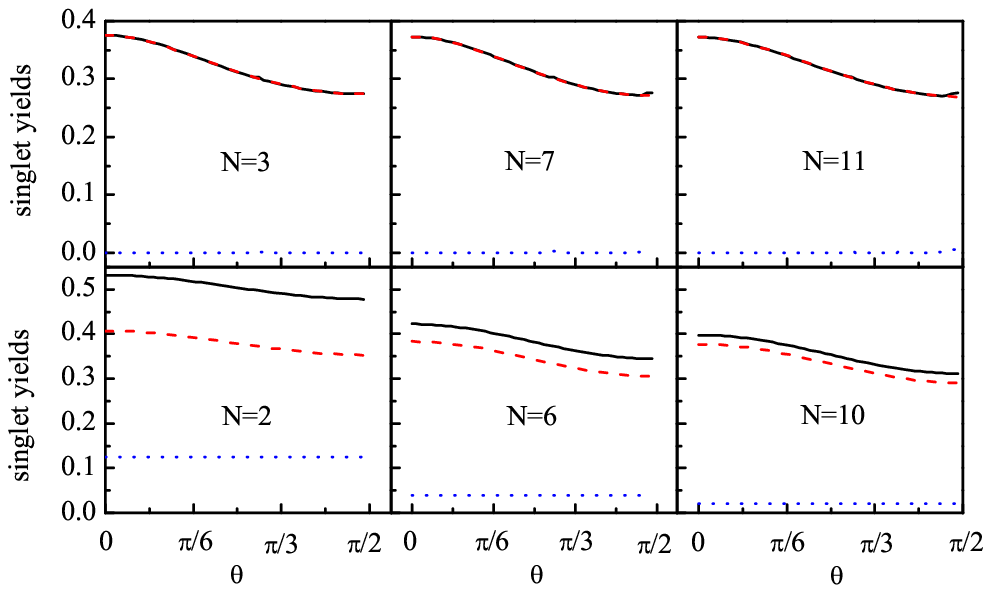}
\parbox{12cm}{\small{Fig. A1} (Color online) The singlet yields $\Phi_{s}$ (black-solid line), $\Phi_{c}$ (blue-dotted line) and $\Phi_{p}$ (red-dashed line) as functions of the direction angle $\theta$ for different $N$. $T_{z}=5\Lambda$, $T_{x}=3\Lambda$. It should be noted that $\Phi_{p}$ for the odd nuclear numbers coincides with $\Phi_{s}$.}
\end{center}
\end{widetext}

Now we consider the horizontal hyperfine interaction, and we still assume that the hyperfine interaction strengths for all the nuclei are the same, i.e., $A_{nx}=A_{ny}=T_{x}/2$ and $A_{nz}=T_{z}/2$ for all $n$. In this case the Hamiltonian (1) in the main text can be expressed as $\hat{H}_{0}=\gamma \textbf{B}\cdot(\hat{S}_{1}+\hat{S}_{2})+T_{x}(\hat{S}_{1+}\hat{J}_{-}+\hat{S}_{1-}\hat{J}_{+})+T_{z}\hat{S}_{1z}\hat{J}_{z}$, where $\hat{S}_{1\pm}=\hat{S}_{1x}\pm i\hat{S}_{1y}$ and $\hat{J}_{\pm}=\sum_{n}\hat{I}_{nx}\pm i\hat{I}_{ny}$. It is convenient that we write the initial nuclear spin state (the completely mixed state) in the angular-momentum representation, i.e., $\rho_{I}(0)=\frac{1}{2^{N}}\sum_{J}\sum_{M}\nu(N,J)|J,M\rangle\langle J,M|$. The action of $\hat{J}_{\pm}$ on $|J,M\rangle$ is given by $J_{\pm}|J,M\rangle=\sqrt{(J\pm M+1)(J\mp M)}|J,M\pm1\rangle$. Setting $T_{x}=3\Lambda$ and $T_{z}=5\Lambda$ as an example, we numerically calculate the singlet yield as shown in Fig. A1. For $N$ being odd, $\Phi_{c}$ is always 0 for different $N$ and $\Phi_{p}$ always coincides with $\Phi_{s}$, which means that the singlet yield is determined completely by the dark state population. But if the total number $N$ is even, $\Phi_{c}$ is always a constant (not zero for less $N$), and $\Phi_{p}$ always differs by a constant from that of $\Phi_{s}$ for all $\theta$, i.e., the magnetic sensitivity $\partial\Phi_{p}/\partial\theta$ for dark state population is the same as $\partial\Phi_{s}/\partial\theta$. The distance between $\Phi_{p}$ and $\Phi_{s}$ decreases with the increasing of $N$. If $N$ is sufficiently large, $\Phi_{c}$ becomes 0 and $\Phi_{p}$ coincides with $\Phi_{s}$. That is to say, the magnetic sensitivity for even nuclear number is also determined completely by the dark state population.

\begin{center}
\includegraphics[width=8cm]{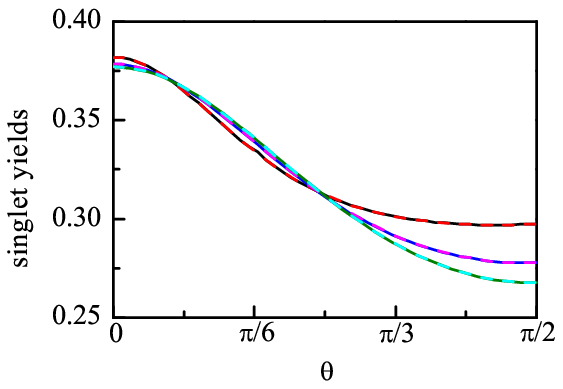}
\parbox{8cm}{\small{Fig. A2} (Color online) The singlet yields $\Phi_{s}$ (solid lines), $\Phi_{p}$ (dashed lines) as functions of the direction angle $\theta$ for different field intensities $B_{0}=32.2\mu\rm T$ (olive-solid and cyan-dashed lines), $B_{0}=46\mu\rm T$ (blue-solid and magenta-dashed lines), $B_{0}=59.8\mu \rm T$ (black-solid and red dashed lines). $A_{x}=A_{y}=3\Lambda$, and $A_{z}=5\Lambda$. It should be noted that $\Phi_{p}$ coincides with $\Phi_{s}$ for different field intensities.}
\end{center}

\section{Two important experimental results}
There are two important experimental results which strongly support the RP model. One is that the birds are able to ``train" to different field strengths: If the field intensity changes in a suitable regime, for example the field intensity is increased or decreased by about $30\%$ of the local geomagnetic field, the birds will disorient temporarily but rework after a sufficiently long time to adapt themselves \cite{Wiltschko72,Wiltschko78,Wiltschko06,Wiltschko13}. The other is that a very weak oscillating field (generally 150nT or even 15nT) whose frequency is resonant with the electron spin Larmor frequency in the geomagnetic field can disorient the birds completely.  But if the oscillatory frequency is detuning from the Larmor frequency, the birds can not be disoriented \cite{Ritz04,Ritz05,Ritz09,Wiltschko15}. Now we investigate the influences of the field intensities and the weak oscillating field on the singlet yield from the point of view of the dark state population one by one. For simplicity, we only consider the single-nucleus RP model.

Firstly we investigate the effect of the field intensity. For relatively strong hyperfine coupling, we can expand $\sin\theta_{\pm}$ and $\cos\theta_{\pm}$ by $\gamma B_{0}/A_{z}$, and $f_{p}(t)$ (see Eq. (17) in the main text) can be expressed as
\begin{equation}\label{fp3}
    f_{p}(t)\approx\frac{1}{2}-\frac{1}{4}\sin^{2}\theta-\frac{\gamma^{2}B^{2}_{0}}{A_{z}^{2}}\biggl(\frac{3}{4}\sin^{2}\theta-\sin^{4}\theta\biggr).
\end{equation}
Substituting Eq. (\ref{fp3}) into Eq. (11) in the main text, one can obtain
\begin{equation}\label{phip3}
  \Phi_{p}\approx\frac{1}{2}-\frac{1}{4}\sin^{2}\theta-\frac{\gamma^{2}B^{2}_{0}}{A_{z}^{2}}\biggl(\frac{3}{4}\sin^{2}\theta-\sin^{4}\theta\biggr).
\end{equation}
It can be seen that the field intensity controls $f_{p}(t)$ (see Eq. (\ref{fp3})), thus affects the singlet yield (see Eq. (\ref{phip3})). The change of the field intensity, for example increasing or decreasing by about $30\%$ of the local geomagnetic field, will induce the change of the singlet yield, so that it disorients the bird transiently. But from Eq. (\ref{phip3}) it can be seen that the singlet yield is mainly determined by $1/2-\sin^{2}\theta/4$ which decreases monotonously with $\theta$. And only the second order of $\gamma B_{0}/A_{z}$ influences $\Phi_{p}$ (or $f_{p}(t)$). Although changing the field intensity will change the singlet yield, the monotonicity will be preserved. This monotonicity preservation ensures that the bird reworks after a sufficiently long time to adapt itself. When we consider the horizontal hyperfine interaction $A_{x}=A_{y}=3\Lambda$ and $A_{z}=5\Lambda$, we numerically calculate the singlet yield for different magnetic fields as shown in Fig. A2. From Fig. A2 we can see that for different fields $\Phi_{p}$ always coincides with $\Phi_{s}$ along different directions, which means that the dark state population determines the singlet yield completely. We also find that $\Phi_{p}$ ($\Phi_{s}$) for different field intensities decreases monotonously with the direction angle and the $30\%$ weaker ($32.2 \mu \rm T$) and stronger ($59.8 \mu \rm T$) fields influence the angular profile evidently. The changes of the angular profile means that the birds will disorient if the field intensity changes suddenly, but the preservation of the monotonicity ensures that the bird can reorient after a long time to adapt itself.

Next we investigate the influence of the weak oscillating field:
\begin{equation}\label{Brf}
   \textbf{B}_{\textbf{rf}}=B_{rf}\cos\omega t(\sin\alpha \cos\beta, \sin\alpha \sin\beta, \cos\alpha),
\end{equation}
where $B_{rf}=150\rm nT$ is the strength of the additional oscillating field with frequency $\omega$, and $\alpha$ and $\beta$ give the direction of the oscillating field. Due to the axial symmetry of the hyperfine interaction tensor we set $\beta=0$, and only focus on $\alpha=\theta+\pi/2$, i.e., the radio frequency field is orthogonal to the geomagnetic field. When we only consider the vertical hyperfine coupling, using the time-dependent perturbation theory \cite{Xu}, we can obtain:
\begin{equation}\label{fprf}
\begin{split}
f^{rf}_{p}(t)&\approx f_{p}(t)-\frac{1}{16}\gamma^{2}B^{2}_{rf}\cos^{2}(\theta_{+}-\theta)t^{2} \\
&-\frac{1}{16}\gamma^{2}B^{2}_{rf}\cos^{2}(\theta_{-}-\theta)t^{2}.
\end{split}
\end{equation}
Substituting Eq. (\ref{fprf}) into Eq. (11) in the main text we can obtain
\begin{equation}\label{phiprf}
    \Phi^{rf}_{p}\approx\Phi_{p}-\frac{\gamma^{2}B^{2}_{rf}}{4k^2}\biggl[\cos^{2}(\theta_{+}-\theta)+\cos^{2}(\theta_{-}-\theta)\biggr].
\end{equation}
It is shown that the weak oscillating field influences the dark state population $f_{p}(t)$ (see Eq. (\ref{fprf})), so that destroys the singlet yield anisotropy, and disorients the bird completely. When we consider the horizontal hyperfine coupling $A_{x}=A_{y}=3\Lambda$ and $A_{z}=5\Lambda$, we numerically calculate the singlet yield under the influence of the oscillating field as shown in Fig. A3. It can be seen that under the influence of the oscillating field, $\Phi_{p}$ coincides with $\Phi_{s}$ and flattens out. $\Phi_{p}$ coincides with $\Phi_{s}$ means that the singlet yield is still completely determined by the dark state population, and the angular profile flattens out means that under the influence of the resonant radio frequency field the bird can not distinguish different directions $\theta$ and disorients completely.

\section{Singlet yield anisotropy comes from the initial state anisotropy}
Recently, Hogben et al. has pointed out that the singlet yield anisotropy (the singlet yield is different for different $\theta$), is essential to the magnetic sensitivity, which not only can come from the anisotropic hyperfine interaction but also can come from the anisotropic initial state \cite{Hannah}. More specifically, if the initial state is isotropic (for example the singlet state $|S\rangle$), the anisotropic hyperfine interaction ($A_{nx}=A_{ny}\neq A_{nz}$) can induce the singlet yield anisotropy; and if the hyperfine interaction is isotropic ($A_{nx}=A_{ny}=A_{nz}$), the anisotropic initial state (for example the triplet state $|T_{0}\rangle=(|10\rangle+|01\rangle)/\sqrt{2}$) can also induce the yield anisotropy. The singlet yield anisotropy induced by the anisotropic hyperfine interaction has been investigated in the main text and in Appendix A. For the case that the singlet yield anisotropy is induced by the anisotropic initial state, whether the singlet yield is still completely determined by the dark state population needs to be discussed.

\begin{center}
\includegraphics[width=8cm]{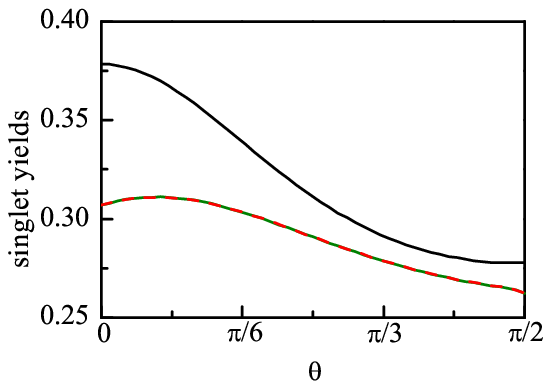}
\parbox{8cm}{\small{Fig. A3} (Color online) The singlet yields $\Phi^{rf}_{s}$ (olive-solid line) and $\Phi^{rf}_{p}$ (red-dashed line) under the influence of the oscillating field compared $\Phi_{s}$ (black-solid line) without considering the radio frequency field. $B_{rf}=150\rm nT$, $\omega=1.315\rm MHz$, $B_{0}=46\mu\rm T$, $A_{x}=A_{y}=3\Lambda$, and $A_{z}=5\Lambda$. It should be noted that under the influence of the oscillating field $\Phi^{rf}_{p}$ coincides with $\Phi^{rf}_{s}$.}
\end{center}

Let two electrons be initially in the triplet state $|T_{0}\rangle=(|10\rangle+|01\rangle)/\sqrt{2}$ and the hyperfine tensors for single-nucleus and multi-nuclei RPs are $A_{x}=A_{y}=A_{z}=5\Lambda$ and $A_{nx}=A_{ny}=A_{nz}=5\Lambda$ for all $n$, respectively. We numerically calculate the singlet yields for different nuclear number and show the results in Fig. A4. As shown in Fig. A4 for the single-nucleus RP, $\Phi_{p}$ always coincides with $\Phi_{s}$ and $\Phi_{c}$ is always zero for different $\theta$. For the multi-nuclei RP model, if the nuclear number is odd, $\Phi_{p}$ is also coincident with $\Phi_{s}$ and $\Phi_{c}$ is always zero. If the nuclear number is even, there is a little difference between $\Phi_{p}$ and $\Phi_{s}$. Through our numerical calculation we find that $\Phi_{p}$ is approaching $\Phi_{s}$ and $\Phi_{c}$ is close to zero as $N$ is increasing. In one word the magnetic sensitivity for the anisotropic initial state and isotropic hyperfine interaction is also determined completely by the dark state population. We also consider other initial states and other hyperfine tensors, and find that the magnetic sensitivity is always determined by the dark state population.

\begin{widetext}
\begin{center}
\includegraphics[width=12cm]{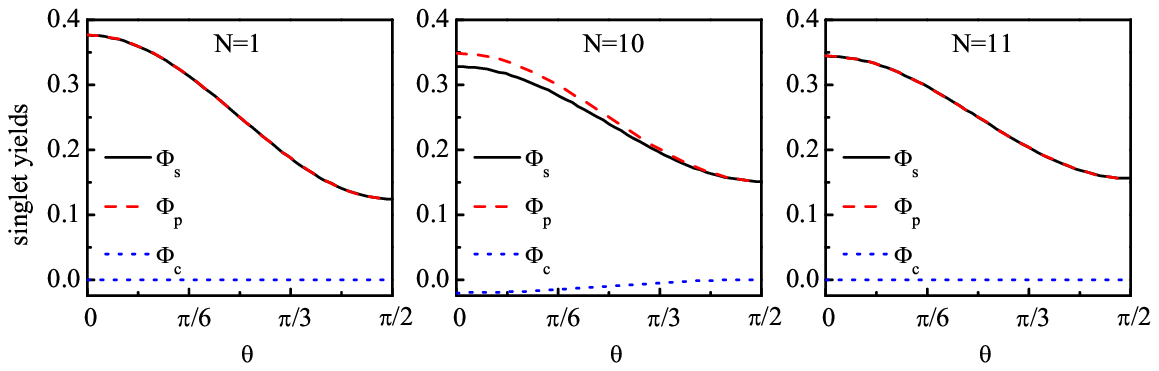}
\parbox{12cm}{\small{Fig. A4} (Color online) The singlet yields as functions of the direction angle $\theta$ for the single-nucleus RP with $A_{x}=A_{y}=A_{z}=5\Lambda$ and the multi-nuclei RP model with $A_{nx}=A_{ny}=A_{nz}=5\Lambda$ for all $n$. Two electrons are initially in the triplet state $|T_{0}\rangle=(|10\rangle+|01\rangle)/\sqrt{2}$. It should be noted that $\Phi_{p}$ for odd nuclear numbers coincide with $\Phi_{s}$.}
\end{center}

\section{Derivation of the master equation}
\label{master equation}
Considering the classical noise, the total Hamiltonian is
\begin{equation}
  H(t)=H_{0}+H'(t)=H_{0}+h(t)\hat{h}.
\end{equation}
In the interaction picture, the Liouville's equation can be written as ($\hbar=1$)
\begin{equation}\label{Liouville's equation}
\frac{d}{dt}\rho_{I}(t)=-i[H_{I}(t),~\rho_{I}(t)],
\end{equation}
where $\rho_{I}(t)=e^{iH_{0}t}\rho(t)e^{-iH_{0}t}$, $H_{I}(t)=e^{iH_{0}t}H'(t)e^{-iH_{0}t}=h(t)e^{iH_{0}t}\hat{h}e^{-iH_{0}t}=h(t)\hat{h}_{I}(t)$.
Generally, Eq. (\ref{Liouville's equation}) can be solved by iteration \cite{Loreti,Wang},
\begin{equation}\label{rhot1}
\begin{split}
\rho_{I}(t)=\rho_{I}(0)-i\int_{0}^{t}dt_{1}h(t_{1})[\hat{h}_{I}(t_{1}),\rho_{I}(0)] -\int_{0}^{t}dt_{1}\int_{0}^{t_{1}}dt_{2}h(t_{1})h(t_{2})[\hat{h}_{I}(t_{1}),[\hat{h}_{I}(t_{2}),\rho_{I}(0)]]+\cdot\cdot\cdot.
\end{split}
\end{equation}
Due to the noise, the ensemble average density matrix satisfies the following equation:
\begin{equation}\label{rhot2}
\begin{split}
\bar{\rho}_{I}(t)=\rho_{I}(0)-i\int_{0}^{t}dt_{1}\langle h(t_{1})\rangle[\hat{h}_{I}(t_{1}),\rho_{I}(0)] -\int_{0}^{t}dt_{1}\int_{0}^{t1}dt_{2}\langle h(t_{1})h(t_{2})\rangle[\hat{h}_{I}(t_{1}),[\hat{h}_{I}(t_{1}),\rho_{I}(0)]]+\cdot\cdot\cdot.
\end{split}
\end{equation}
We consider a Gaussian white noise, i.e., $\langle h(t)\rangle=0$, thus the $n'$th-order correlation can be written as
\begin{equation}
\begin{split}
\langle h(t_{1})h(t_{2})\cdot\cdot\cdot h(t_{n})\rangle=
\begin{cases}
0& \text{if $n$ is odd},\\
\sum\limits_{\begin{subarray}{l}
all~(n-1)!!\\ ~pairings
\end{subarray}}\langle h(t_{1})h(t_{2})\rangle\langle h(t_{3})B'(t_{4})\rangle\cdot\cdot\cdot\langle h(t_{n-1})h(t_{n})\rangle& \text{if $n$ is even},
\end{cases}
\end{split}
\end{equation}
with $(n-1)!!=(n-1)(n-3)\cdot\cdot\cdot5\cdot3\cdot1$ \cite{Wolf}.
We assume that $\langle h(t)h(\tau)\rangle=\Gamma\delta(t-\tau)$, i.e., the Markovian process, and obtain
\begin{equation}\label{rhot3}
     \bar{\rho}_{I}(t) =\rho_{I}(0)
    -\int_{0}^{t}dt_{1}\Gamma[\hat{h}_{I}(t_{1}),[\hat{h}_{I}(t_{1}),~\rho_{I}(0)]]
    +\int_{0}^{t}dt_{1}\Gamma^{2}[\hat{h}_{I}(t_{1}),[\hat{h}_{I}(t_{1}),\int_{0}^{t_{1}}dt_{2}[\hat{h}_{I}(t_{2}),[\hat{h}_{I}(t_{2}),\rho_{I}(0)]]]]+\cdot\cdot\cdot, \\
\end{equation}

which is just the iterative expression of the following differential equation \cite{Loreti,Wang},
\begin{equation}
   \frac{d}{dt}\bar{\rho}_{I}(t)=-\Gamma[\hat{h}_{I}(t),[\hat{h}_{I}(t),\bar{\rho}_{I}(t)]].
\end{equation}
In the Schr\"{o}dinger picture, it can be written as
\begin{equation}
\frac{d}{dt}\bar{\rho}(t)=-i[H_{0},\bar{\rho}(t)]-\Gamma[\hat{h},[\hat{h},\bar{\rho}(t)]].
\end{equation}
\end{widetext}

\end{document}